\documentclass[letterpaper, 10 pt, conference]{ieeeconf}  

\usepackage{cite}
\usepackage{amsmath,amsfonts}
\usepackage{graphicx}
\usepackage{textcomp}
\usepackage{xcolor}

\usepackage{booktabs} 
\usepackage{verbatim}
\usepackage{graphics} 
\usepackage{epsfig} 
\usepackage{soul}
\usepackage{epstopdf}
\usepackage{subfigure}
\usepackage{balance}
\usepackage{comment}
\usepackage{booktabs}
\usepackage{amsmath}
\usepackage{amsfonts}
\usepackage{bm}
\usepackage{colortbl}
\usepackage{tabularx}
\usepackage{color}
\usepackage{xspace}
\usepackage{graphicx}    
\usepackage{url}         
\usepackage{algorithm}
\usepackage{algpseudocode}
\usepackage{placeins}
\usepackage{multirow}
\let\labelindent\relax
\usepackage{enumitem}
\usepackage{leading}
\usepackage{pgfplots}
\usepackage{tikz}
\usepackage{xcolor}

\usepackage{amsmath}

\IEEEoverridecommandlockouts                              

\title{\LARGE \bf
rWiFiSLAM: Effective WiFi Ranging based SLAM System in Ambient Environments
}

\author{Bo Wei$^{1*}$, Mingcen Gao$^{2}$, Chengwen Luo$^{3}$, Sen Wang$^{4}$, Jin Zhang$^{3}$
\thanks{* Corresponding Author}
\thanks{$^{1}$Bo Wei is with School of Computing and Communications,
        Lancaster University, Lancaster, UK.
        {\tt\small bo.wei@lancaster.ac.uk}}%
\thanks{$^{2}$Mingcen Gao is with Google, San Francisco, USA
        {\tt\small gaomingcen@gmail.com}}%
\thanks{$^{3}$Chengwen Luo and Jin Zhang are with College of Computer Science and Software Engineering, Shenzhen University, Shenzhen, China
{\tt\small chengwen@szu.edu.cn, jin.zhang@szu.edu.cn} }%
\thanks{$^{3}$Sen Wang is with Department of Electrical and Electronic Engineering, Imperial College London, London, UK
{\tt\small sen.wang@imperial.ac.uk} }%
\thanks{This paper is funded by Lancaster University Start-up Funding and Lancaster Security. }%
}

\begin{document}

\maketitle
\thispagestyle{empty}
\pagestyle{empty}


\begin{abstract}
In this paper, we propose rWiFiSLAM, an indoor localisation system based on WiFi ranging measurements. Indoor localisation techniques play an important role in mobile robots when they cannot access good quality GPS signals in indoor environments. Indoor localisation also has many other applications, such as rescue,  smart buildings, etc. Inertial Measurement Units (IMU) have been used for Pedestrian Dead Reckoning (PDR) to provide localisation services in the indoor environment as it does not rely on any other signals. Although PDR is a promising technique, it still suffers from unavoidable noise and bias from IMUs in mobile devices. Loop closure is necessary for these scenarios. In this paper, we design an efficient loop closure mechanism based on WiFi ranging measurements along with IMU measurements in a robust pose graph SLAM framework for indoor localisation. One novelty of the proposed method is that we remove the requirement of the full knowledge of the WiFi access point locations, which makes our proposed method feasible for new and/or dynamic environments. We evaluate our designed system in real environments and show the proposed method can achieve sub-meter localisation accuracy and improve the localisation performance by more than 90\% compared with the IMU based PDR. 

\end{abstract}

\section{INTRODUCTION}\label{sec:intro}
Localisation is a fundamental research topic with a wide range of applications, such as mobile robotics, self-driving cars, and smart buildings. Localisation also plays an important role in mobile systems, such as mobile phones, to support many services. Global Positioning System (GPS) is a common localisation solution, where devices with GPS receive signals from satellites to localise themselves. However, due to the weak GPS signal perceived in the indoor environment, the GPS localisation accuracy significantly drops, which cannot provide sufficiently good performance.

\begin{figure}[!t]
   \centering
    \includegraphics[width=.3\textwidth]{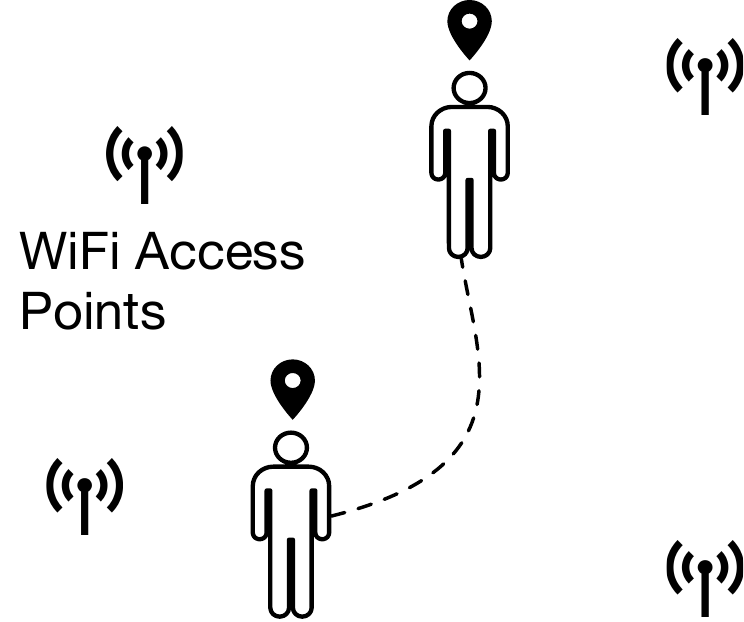}
    \caption{Example of the proposed system: using IMU and WiFi RTT measurements for indoor localisation}\label{fig:example}
\end{figure}

To solve these challenges, many indoor localisation solutions have been designed for mobile devices \cite{Jones2011,Huang:2011ez,roetenberg2009xsens}. Inertial Measurement Units (IMU) are common sensors integrated into mobile devices as motion sensors and leveraged for Pedestrian Dead Reckoning (PDR). Although it can provide promising localisation performance, the IMU sensor bias generates unavoidable accumulated trajectory drift. In order to correct the drift in IMU-based PDR solutions, supplementary sensors, such as cameras \cite{Jones2011}, mmWave radar \cite{lu2020see}, etc., are used. Since cameras and mmWave radar can capture high-quality and high-resolution features of the surrounding environment, they can achieve good and stable localisation performance. However, cameras and mmWave radar cost a high volume of energy, which is not suitable for long-term usage on mobile devices. Other radio signal based solutions have been designed, such as Ultra-wide Band (UWB) based \cite{roetenberg2009xsens} and magneto-inductive device based \cite{wei2018imag}, but they are not commonly equipped within mobile devices. 

WiFi, as a common communication technique, has been widely integrated into mobile systems. Received Signal Strength Indicator (RSSI) from WiFi has been used with IMU sensors for indoor localisation, such as \cite{Huang:2011ez,Ferris2007,Mirowski2013}. The localisation performances of these systems are promising, but they need heavy signal processing methods to mitigate the WiFi signal variance caused by multi-path effect \cite{wei2015drti}. Although RSSI has a certain relationship with the distance between a device and a WiFi access point, the RSSI variance in a certain location is too huge for accurate localisation. This is because RSSI was not designed for localisation purposes, and multi-path effect is very common in the indoor environment. Recently, a new protocol IEEE 802.11mc with WiFi Round Trip Time (RTT) feature has been designed for ranging between an RTT capable WiFi access point and a mobile device \cite{ieee80211,android80211}. With the deployment of multiple APs with the WiFI RTT (at least 3 access points for 2D localisation and 4 access points for 3D localisation), a mobile system can locate itself in the area of interest. However, since trilateration is usually used on the ranging results between a mobile device and APs, the exact locations of APs are required. This condition requires frequent reconfiguration and the information exchanged in a new environment or a dynamic environment, such as when WiFi routers are moved. The configuration, i.e., surveying and updating the map of WiFi access points, is labour expensive, especially in a large and/or dynamic environment, not to mention localisation in an uncooperative environment. Additionally, due to multiple path effects in an indoor environment \cite{ibrahim2018verification}, current localisation solutions only relying on WiFI RTT ranging results may decrease the indoor localisation performance.

In this paper, we propose a SLAM system based on data fusion on WiFi RTT and inertial measurements, which achieves good accuracy in indoor localisation.  
Different from existing research, we do not need the prior knowledge of the locations of WiFi access points and the dependence on accurate ranging results. Instead, the ranging measurements are used as \textbf{observations} directly for clustering. As discussed, the IMU only based PDR can provide an accurate trajectory in a relatively short term, but with unavoidable long term drift. In our paper, we propose to use clusters from WiFi ranging measurements for loop closure and the PDR drift correction. One novelty of the system is the removal of the configuration requirement of the deployment of WiFi access points, which enables the proposed system suitable for a dynamic environment, such as home environment, where APs are not fixed and/or easily moved, or a new environment, where mobile devices do not contain the location information of WiFi access points. Furthermore, we consider the false positives of loop closures and use a robust pose graph SLAM framework to provide high quality indoor localisation performance.

To summarise, the contributions of this paper are as follows:
 \begin{itemize}
     \item We design an indoor localisation system based on the observations of WiFi ranging measurements within a pose graph SLAM framework. Our designed system removes the prior knowledge of the locations of WiFi access points and the dependence on accurate ranging results. 
     \item We design an effective clustering method for loop closure along with a robust SLAM framework to tolerate false positives of loop closing constraints and provide accurate localisation performance. 
     \item We have conducted extensive experiments in a real environment, which shows our proposed method has good localisation performance. 
 \end{itemize}

The rest part of this paper is organised as follows. Section \ref{sec:related} discusses the related works, and we show the overview of the designed system in Section \ref{sec:system}. Section \ref{sec:method} details the proposed accurate indoor localisation system. Next, in Section \ref{sec:exp}, extensive experiments are performed in a real environment, which shows a good and robust localisation performance of the proposed system. Finally, we conclude the paper in Section \ref{sec:conclusion}.
 
\section{RELATED WORKS}\label{sec:related}
In this section, we will discuss the related works regarding indoor localisation systems. 
As a ubiquitous signal, WiFi has been studied for indoor localisation.  \cite{Ferris2007,Huang:2011ez,liu2019collaborative} used WiFi received signal strength along with IMUs for indoor localisation. However, due to the multipath effect \cite{wei2015drti}, the use of received signal strength cannot maintain stable quality for indoor localisation. The wireless communication research community has also been aware of the importance of WiFi for indoor localisation, standard IEEE 802.11mc has been designed for ranging \cite{ieee80211} and WiFi routers and mobile devices have started to integrate the new standard \cite{android80211}. The common methods using WiFi RTT measurements are based on multilateration, i.e., localisation based on estimate distances between the mobile device and WiFi access points. The data fusion of WiFi RTT and IMU measurements has also been studied \cite{guo2022robust,gentner2021wifi}. Different from it, our proposed method does not rely on accurate ranging estimate from WiFi RTT measurements. Because the multiple path effect still exists, albeit having better performance compared with the use of received strength, it is not easy to estimate the ranging quality of WiFi RTT. Instead, we focus on the localisation of the mobile device rather than other factors, such as the location estimate of WiFi access points. We use WiFi RTT measurements directly as observations for loop closure and apply the pose GraphSLAM to obtain an accurate trajectory. 

Other wireless signals have also been used for localisation. Bluetooth, as another common wireless communication technique, was also employed to indicate the locations \cite{tekler2020scalable}. However, due to the short communication range of Bluetooth, a large volume of Bluetooth anchor points are needed for localisation purposes. UWB is another wireless communication technique. As UWB is designed for indoor localisation, it can achieve centimetre-level accuracy \cite{decawave}. Along with IMU measurements, it can deliver a high-precision context awareness system, including localisation\cite{roetenberg2009xsens}. The basic principle of UWB for localisation is also multilateration based on the ranging between mobile devices and WiFi access points. One fallback for UWB based applications is that UWB is not as common as WiFi devices, which needs specialised devices and extra installation. 


Additional to wireless technologies, other sensors with high resolutions are used for localisation, such as laser based \cite{Hesch2010}, mmWave radar based \cite{lu2020see}, camera based \cite{Jones2011}. Although they can provide very high accuracy localisation performance, their high energy consumption is one constraint for the large-scale application in mobile devices. Researchers have also investigated other signals for localisation purposes, such as magnetic field \cite{wei2018imag, wei2021imag, wang2016keyframe}, photovoltage \cite{wei2020solarslam}, etc.

There also exist radio based device free localisation methods, i.e., users do not need to carry any devices and the system localises the user in the area of interest based on the radio interference by human beings \cite{wei2015drti}. 

Localisation is also the core technology for self-driving cars \cite{ma2019exploiting}, which uses multiple sensors, such as GPS, IMU, Lidar, and cameras. With the assistance of a powerful computer, self-driving cars can achieve good localisation performance. Different from localisation techniques used in self-driving cars, our paper focuses on the use of readily available sensors on mobile phones, i.e., WiFi and IMU, for indoor localisation.

\begin{figure}[!t]
\vspace{2mm}	
   \centering
    \includegraphics[width=.5\textwidth]{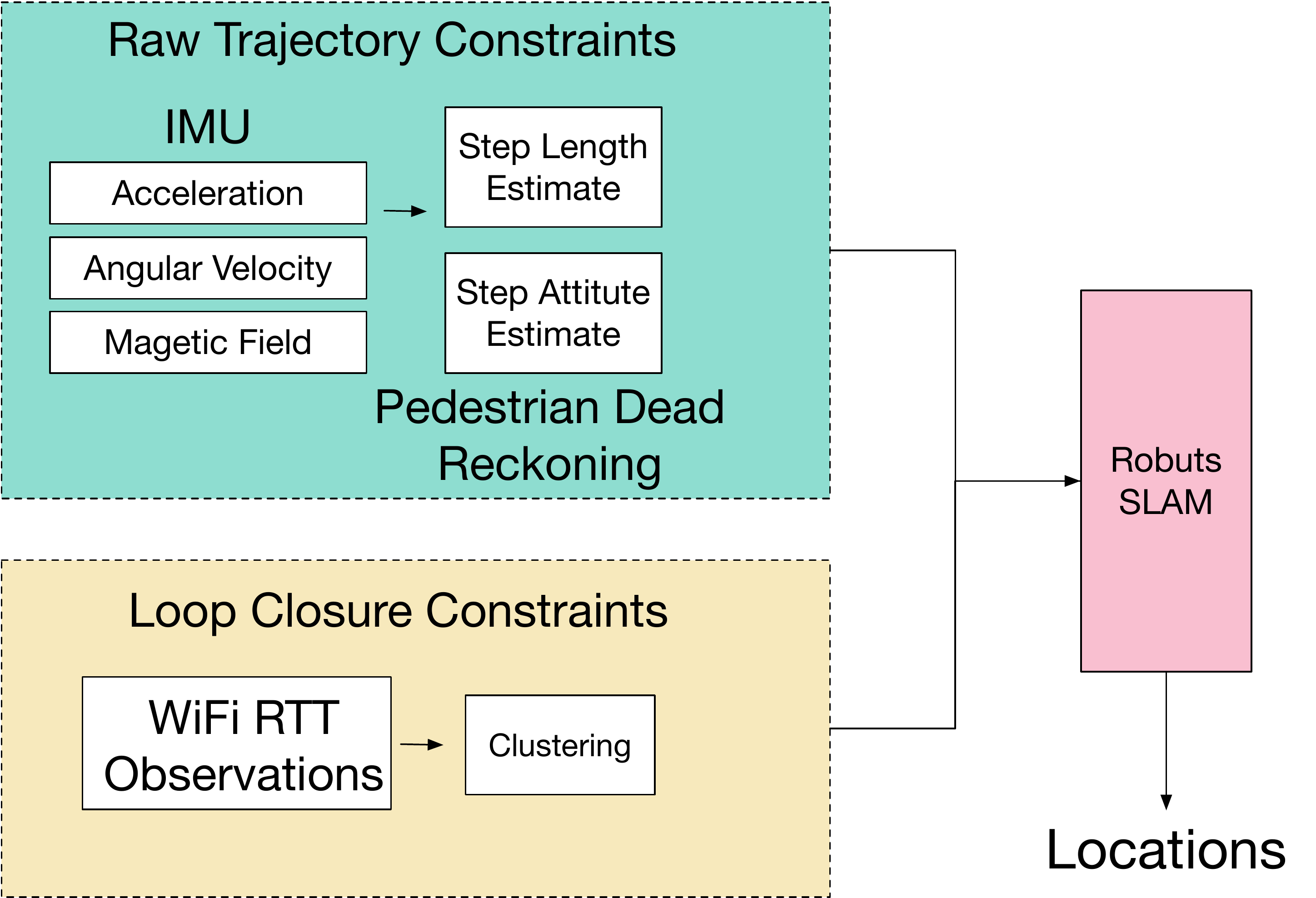}
    \caption{System overview of the proposed indoor localisation system: rWiFiSLAM}\label{fig:overview}
\end{figure}

\section{SYSTEM OVERVIEW}\label{sec:system}
In this section, we overview the whole structure of the system rWiFiSLAM, and the flowchart is shown in Figure \ref{fig:overview}. The user, carrying a mobile phone, walks in an area of interest with the deployment of WiFi access points. The mobile phone, equipped with our designed system, can localise itself. To enable the indoor localisation system, the mobile system continuously collects acceleration, angular velocity measurements and magnetic field measurements from its own IMU, which provides an initial trajectory estimate. WiFi RTT measurements are collected simultaneously, and a clustering method will be applied on WiFi RTT measurements for the loop closure detection and adjusting the biased initial trajectory. A robust pose graph SLAM will then be applied on both initial trajectory and WiFi RTT constraints, which provides an accurate indoor localisation performance. 
\section{METHOD}\label{sec:method}
In this section, we will show the details of the proposed indoor localisation method, which includes Pedestrian Dead Reckoning, an efficient loop closure method based on WiFi RTT measurements and a robust SLAM. 

\subsection{Pedestrian Dead Reckoning}
The first step of the system is to provide an initial trajectory based on inertial navigation. 
A robust Pedestrian Dead Reckoning (PDR) method is firstly applied on the collected acceleration, angular velocity and magnetic field measurements from the accelerometer, gyroscope and magnetometer. Theoretically, an IMU in a mobile device, containing an accelerometer, a gyroscope and a magnetometer, can estimate the carrier's trajectory based on acceleration and gyroscope. However, the low specification IMU in mobile devices cannot provide sufficiently accurate measurements because of the noise in the accelerometer and long term drift in the gyroscope. To solve this problem, the whole trajectory estimate is split into multiple step motion estimates. This is based on the fact that the velocity is 0 when both feet are on the ground during walking and the long term bias can be mitigated at a certain level by dividing the long term estimate into smaller chunks. Therefore, the PDR can be conducted by the following two steps, (1) performing step detection and (2) estimating the length $L$ and attitude $\alpha$ of each step. In our system, acceleration magnitudes are used for step detection. To be specific, the norm of $i$th 3D acceleration measurement $(a_{xi}, a_{yi}, a_{zi})$ is calculated as $a_i = (a_{xi}^2+a_{yi}^2+a_{zi}^2)^{\frac{1}{2}}$. A threshold of standard deviations within a window size $w$ is set to detect the step (relatively steady and the corresponding standard deviation is below the threshold) and moving (the corresponding standard deviation is above the threshold). Once the steps are detected, the step length is estimated using Weinberg algorithm \cite{weinberg2002using} by using vertical movement of the hip. It is empirically shown in \cite{weinberg2002using, ruiz2017pedestrian} that the $j$th step length can be calculated in Equation \ref{equ:step}.
\begin{equation}\label{equ:step}
    L_j = h \times ( \text{max}(\widetilde{a}(j)) - \text{min}(\widetilde{a}(j)) )^{1/4}
\end{equation}
,where $a(j)$ represents the vectors of the magnitude of acceleration measurements (i.e., the norm of 3D acceleration measurements) between $j$th step and ($j$-1)th step after a low pass filter to remove the measurement noise. The maximum and minimal values of this vector are obtained by using max() and min(), respectively. $h$ is a constant. In the meanwhile, the attitude $\alpha$ of each step is estimated by the angular velocity in the global reference. Please note that the raw data from IMUs are based on the IMU reference system. Therefore, the measurements of angular velocity from the IMU reference system need to be transformed to the global reference with the assistance of the magnetometer \footnote{More details of the reference transformation can be found in \cite{woodman2007introduction}.}.  
\begin{figure}[!t]
\vspace{2mm}	
   \centering
    \includegraphics[width=.3\textwidth]{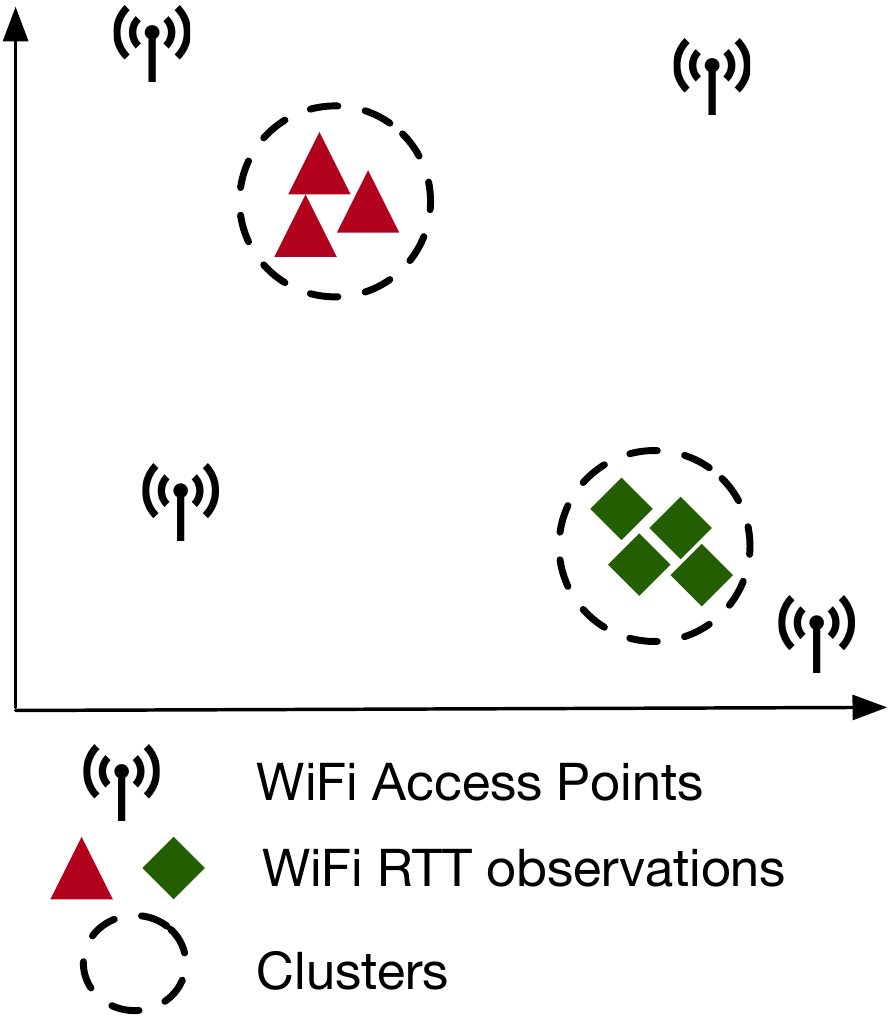}
    \caption{Examples of clustering for loop closures}\label{fig:cluster}
\end{figure}

\begin{figure*}[!t]
\vspace{2mm}	
   \centering
    \includegraphics[width=.9\textwidth]{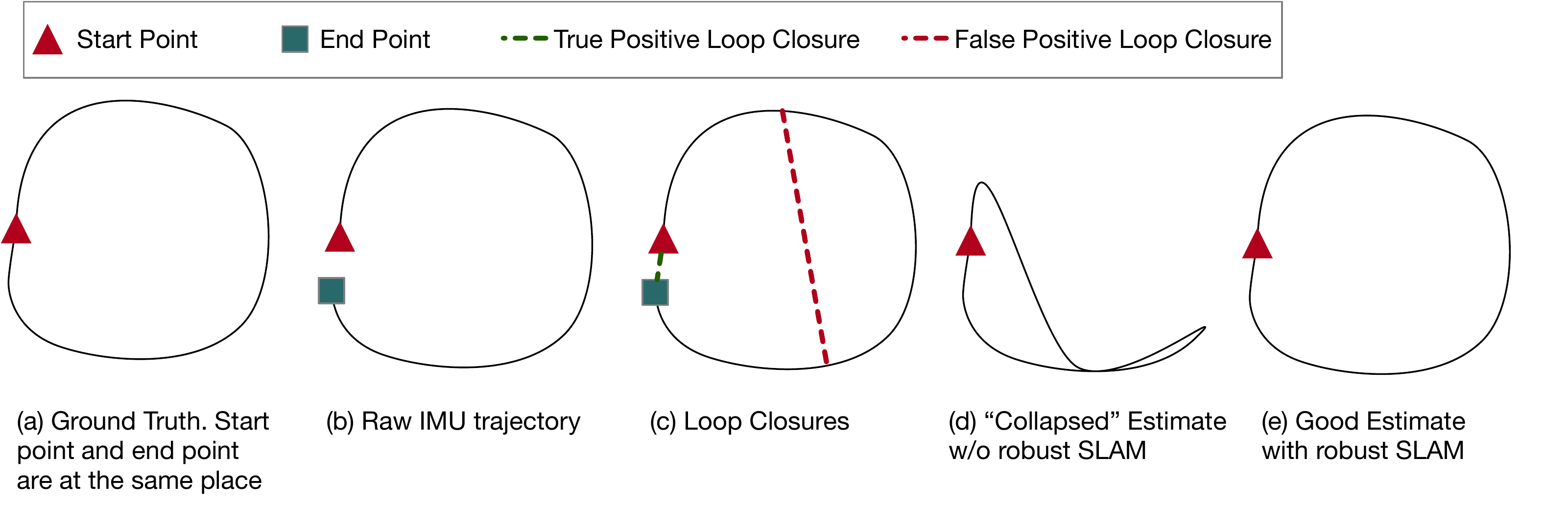}
    \caption{Examples of the use of robust SLAM for localisation: (a) Ground truth of trajectory in which user walks in a cycle; (b) Raw trajectory estimate using IMU is biased; (c) Two loop closures detected. One is a true positive, and the other one is a false positive; (d) The estimated trajectory is ``collapsed'' when using traditional SLAM; (e) Good trajectory estimate when using robust SLAM }\label{fig:slam}
\end{figure*}

\subsection{Efficient WiFi RTT Measurement Clustering for Loop Closure}

Along with trajectory from the PDR method, WiFi RTT measurements are used as observations directly for the loop closure. WiFi RTT was initially designed for indoor localisation based on its ranging capability between a mobile device and a WiFi access point. The common solution using WiFi RTT measurements is trilateration, i.e., the location estimation based on the locations of WiFi access points and the corresponding WiFi RTT ranging measurements. However, due to multipath effect, the ranging from WiFi RTT may not be sufficiently accurate \cite{ibrahim2018verification}.
Different from existing solutions, we do not rely on the knowledge of locations of WiFi access points or accurate ranging results. Instead, we use the WiFi RTT measurements as observations to directly identify loop closures, based on the fact that the mobile device can perceive similar WiFi RTT measurements at the same place, as shown in Figure \ref{fig:cluster}. These similar observations will be used to indicate loop closures, i.e., the user revisits the same place.

Since the locations of WiFi access points are not known in our setting, it is impossible to calculate the exact locations based on WiFi RTT ranging measurements. Therefore, we cannot estimate the relevant locations between two observations except if they are in the vicinity. When a mobile phone revisits the same place, the observations from all WiFi RTT measurements are similar. Our loop closure methods are based on the similarity of the observations containing all the measurements from WiFi access points. 

To enable loop closure, we design an efficient clustering method for WiFi RTT observations. The $i$th WiFi RTT observation is denoted as $w_i = (w_{1i},w_{2i},...,w_{ki},...,w_{ni})$, where $n$ is the number of WiFi access points and $w_{ki}$ is the WiFi RTT measurement between the mobile device and the $k$th WiFi access point in this $i$th observation. Once the $i$th WiFi RTT observation $w_i$ is collected, it is used for clustering with previous WiFi RTT observations, i.e., $\{w_1, w_2, ..., w_{i-1}\}$, by using a Euclidean distance $d_{i,k}  = ||w_i - w_k ||_2$. When $d_{i,k}$ is below a threshold, the corresponding points when $i$th and $k$th WiFi RTT measurements in the raw estimate trajectory are labelled as a constraint for loop closure, which means they are candidates of places in the vicinity. However, due to the multipath effect, WiFi RTT measurements are not necessarily accurate, and many false positives of loop closures exist during the clustering method. To solve this, we use a robust SLAM framework to achieve global optimisation.

\subsection{Robust SLAM Framework}

In this section, we demonstrate the used pose graph SLAM framework in detail. The traditional pose  graph SLAM \cite{Thrun05,grisetti2010hierarchical} conducts optimisation based on all the constraints equally by solving the following optimisation problem in Equation \ref{equ:slam}.
\begin{equation} \label{equ:slam}
\operatorname*{argmin}_A  \sum_{i \in L} r_i^T  M_i r_i
\end{equation}
, where $A$ is the estimated trajectory for the localisation purpose, and $L$ is the constraint set, including IMU raw trajectory and loop closures. 
$M$ is the information matrix of all constraints, and $r$ represents the corresponding error terms. When applying the traditional SLAM framework to our application, the false positives of constraints from WiFi RTT observations are equally treated, which will lead to unnecessary loop closures and decrease indoor localisation performance significantly. In the example in Figure \ref{equ:slam}(d), the estimated trajectory is collapsed due to the false positive of loop closures. 

The robust pose graph SLAM optimiser solves this problem by introducing scaling factors for loop closures. Equation \ref{equ_slam_robust} shows the optimiser of the used robust pose graph SLAM.
\begin{equation}\label{equ_slam_robust}
\operatorname*{argmin}_A   \underbrace{ \sum_{i \in L_{traj}} r_i^T M_i r_i }_{\text{(Trajectory Constraints)}} +  \underbrace{   \sum_{i \in L_{loop}}s_i^2 r_i^T M_i r_i }_{\text{(Loop Closures)}}
\end{equation}
When using the robust pose graph SLAM optimiser, constraints from raw IMU based trajectories and loop closures. Constraints from raw IMU trajectories are used as that in the traditional graph SLAM, while the consideration of loop closures involves a scaling factor $s$ defined in \ref{equ:scalor}. 
\begin{equation} \label{equ:scalor}
    s_i = \operatorname*{min}(1, \frac{2C}{C +  \Theta_i^2} )
\end{equation}
, where $C$ is a free parameter, and $\Theta$ is also the error term of loop constraints \cite{agarwal2013robust}. The error terms for loop closures depend on the used threshold for the loop closure detection. The use of the scaling factor $s$ can eliminate the impact of false positives of loop closure by providing less weight to them and relying more on the constraints from IMU based trajectories when conducting the optimisation. Please refer to \cite{Thrun05,grisetti2010hierarchical,Kummerle2011,agarwal2013robust} for more technical details regarding the traditional pose graph SLAM and robust pose graph SLAM.

\begin{figure}[!t]
\vspace{2mm}
	\centering
	\subfigure[Google Pixel 6 Mobile Phone]{
		\includegraphics[width=0.3\textwidth]{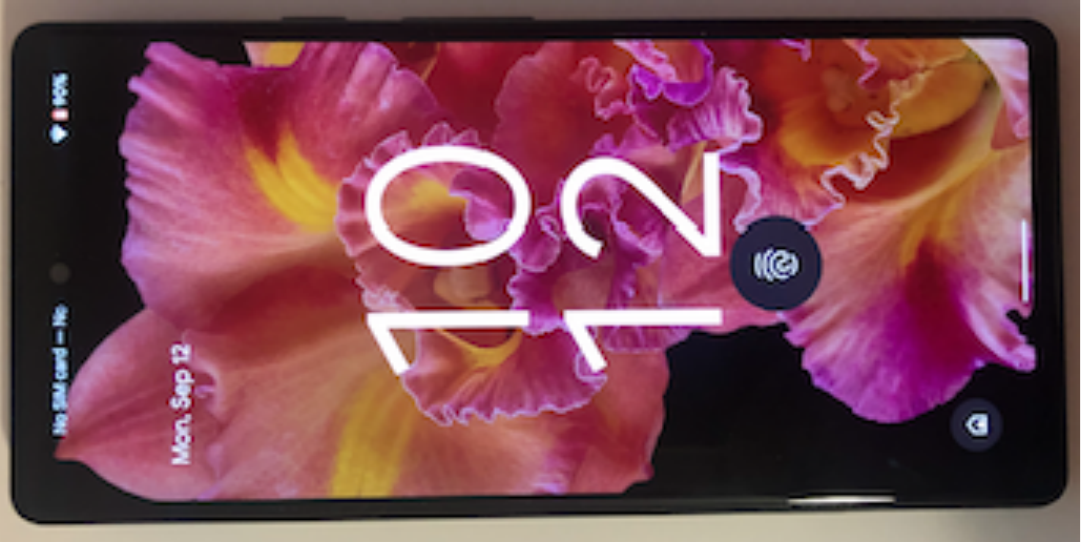}
		\label{fig:hardware_phone}
		}
  
		\subfigure[WiFi Access Points]{
		\includegraphics[width=0.27\textwidth]{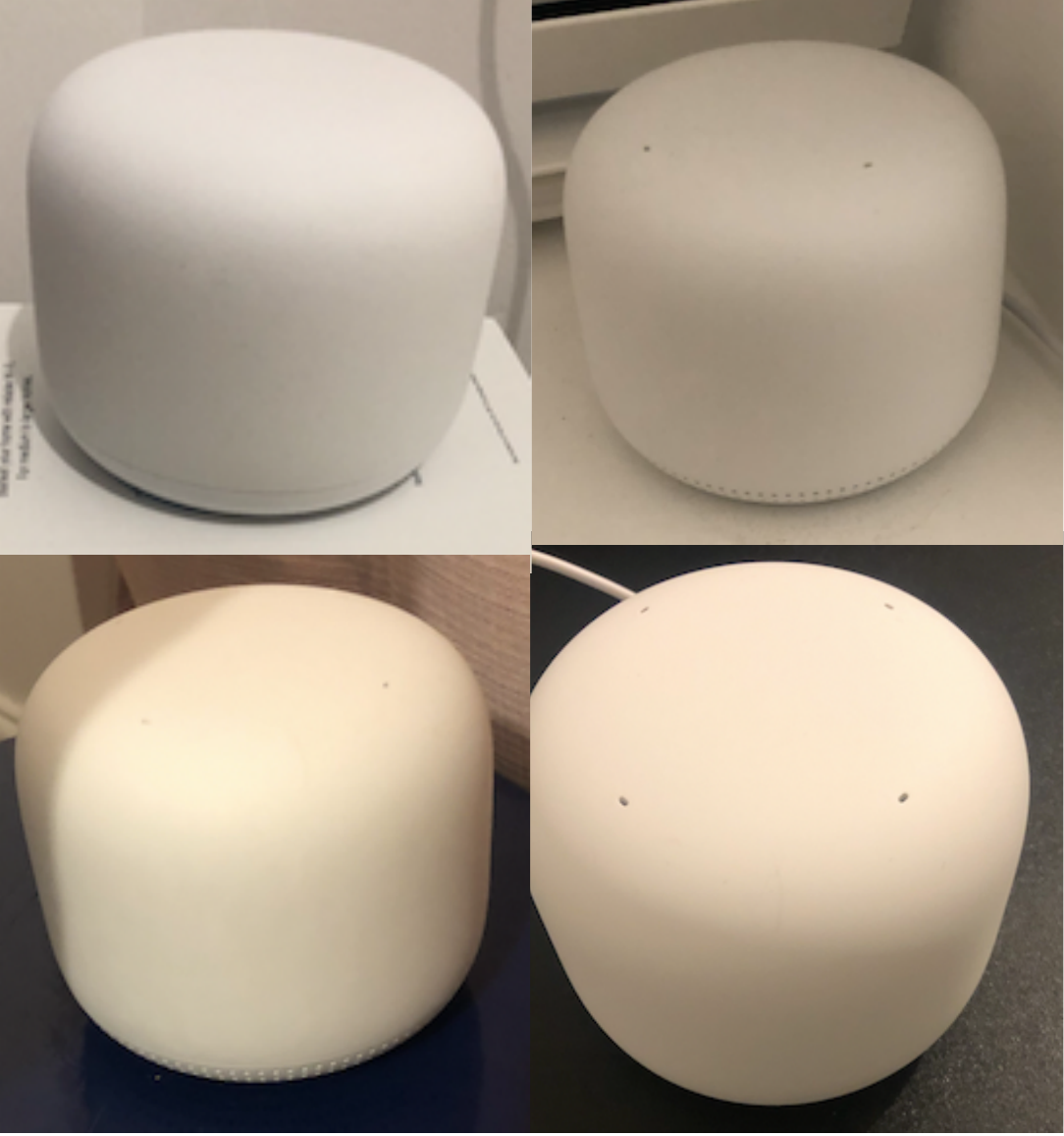}
		\label{fig:hardware_wifi}
		}

		\caption{Hardware: (a) Google Pixel 6 Mobile Phone carried by the user in the designed system. IMU measurements and WiFi RTT measurements are collected using this mobile phone for localisation purpose. }
		\label{fig:hardware}
\end{figure}

\begin{figure*}[!t]
\vspace{2mm}
	\centering
	\subfigure[Ground Truth]{
		\includegraphics[width=0.31\textwidth]{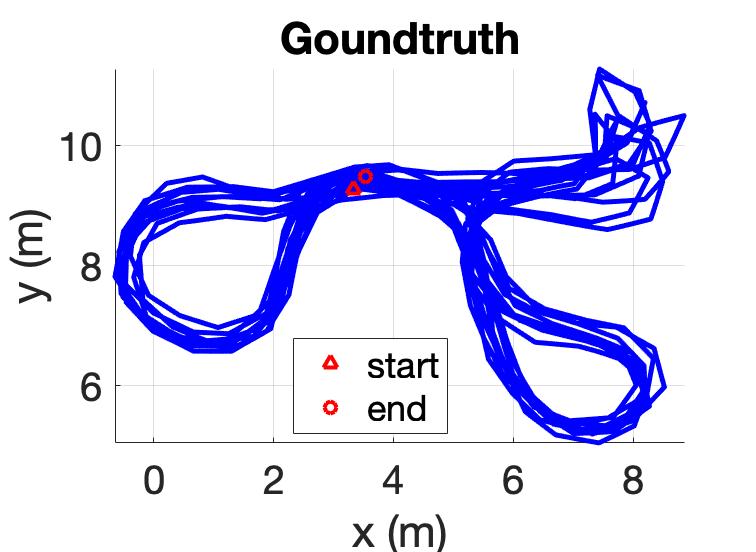}
		\label{fig:traj_gt}
		}
        \subfigure[Trajectory Only Using IMU]{
		\includegraphics[width=.31\textwidth]{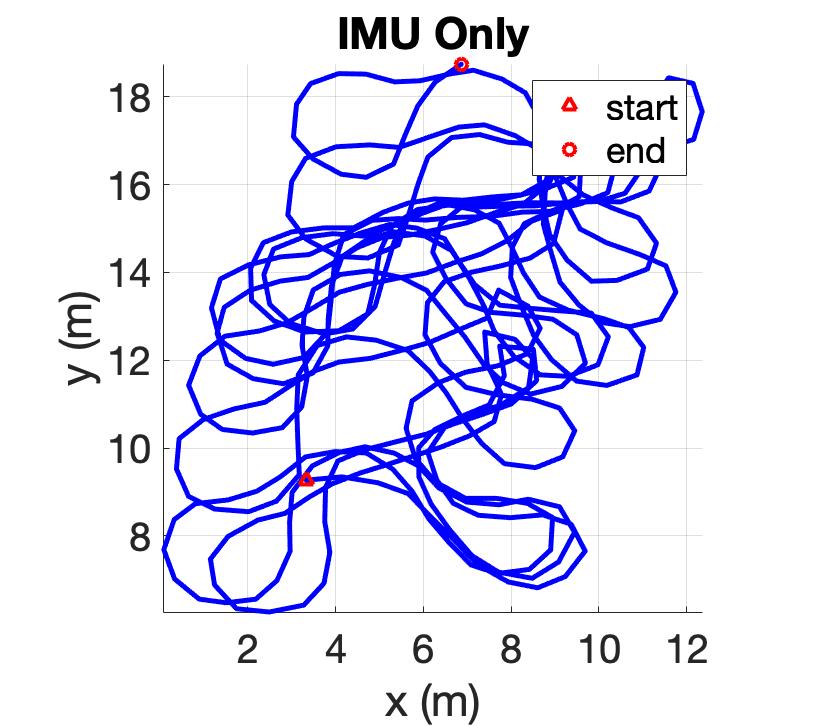}
		\label{fig:traj_imu}
		}
        \subfigure[Trajectory with Traditional SLAM]{
		\includegraphics[width=.31\textwidth]{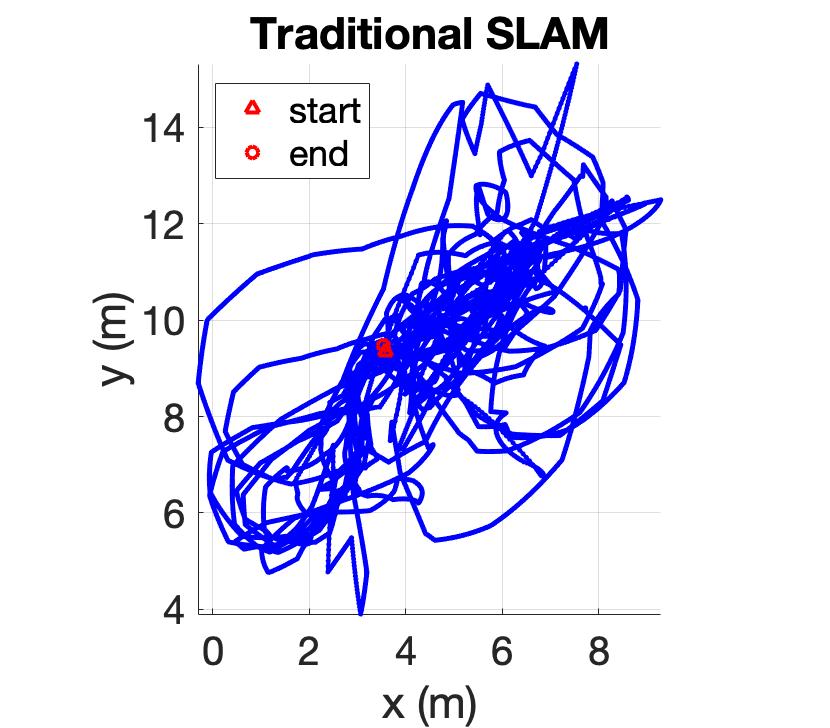}
		\label{fig:traj_noslam}
		}

		\subfigure[Trajectory Using Proposed Method]{
		\includegraphics[width=0.32\textwidth]{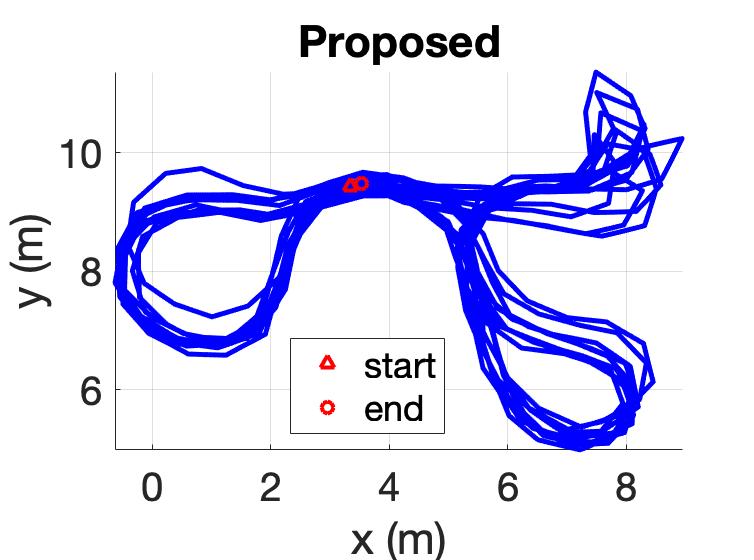}
		\label{fig:traj_slam}
		}
            \subfigure[CDF of localisation errors]{
		\includegraphics[width=.32\textwidth]{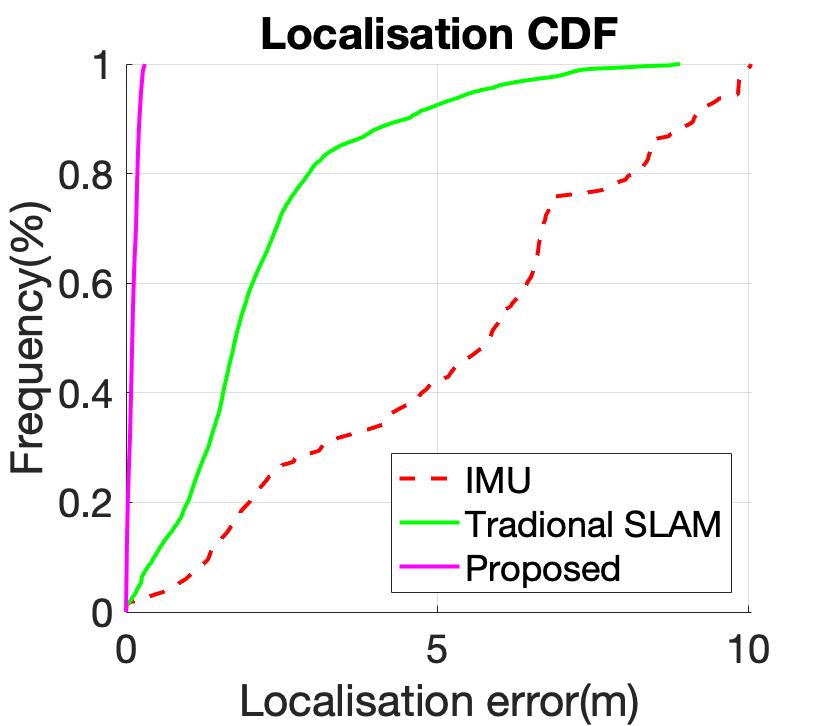}
		\label{fig:cdf}
		}

		\caption{Performances:  (a) Ground Truth; (b) Trajectory Only Using IMU; (c) Trajectory with Traditional SLAM; (d) Trajectory Using Proposed Method; (e) CDF of localisation errors  }
		\label{fig:traj}
\end{figure*}

\section{EVALUATION}\label{sec:exp}
In this section, we will show the details of the used hardware, performance metric and the evaluation of the proposed system in real environments. 
\subsection{Hardware}

Figure \ref{fig:hardware} shows the hardware used in the evaluation. We use one Google Nest WiFi router and three Google Nest WiFi points \cite{googlewifi} as the WiFi access points. The mobile device we use is Google Pixel 6. Other devices supporting IEEE 802.11mc can also be used for the same purpose. The full list of devices supporting IEEE 802.11mc can be found in \cite{android80211}. WiFi access points are deployed on one floor of a house.
There is no specific configuration for Google Nest WiFi routers and Nest WiFi points other than enabling WiFi RTT functions using the official Google WiFi App \cite{WifiRttScan}. For Google Pixel 6, the sampling rate of the IMU is set as 100 Hz, and the sampling rate of WiFi RTT is empirically set as 5 Hz. A higher sampling rate could be set in the mobile device, but that will lead to packet loss because of the burden of processing the ranging requests in the router end\footnote{One WiFi ranging measurement is averaged over a sequence of the ranging requests in the lower level, but the API to access each measurement is not available in the current Android Version Android 12 (version 8.4). }.

\subsection{Performance Metric}
Our proposed system will focus on 2D localisation in the indoor environment.
In this section, we use the following metrics to evaluate the proposed indoor localisation system. 
 \begin{itemize}
     \item \textbf{Trajectories}: Trajectories using the proposed method, IMU based PDR and that using traditional SLAM \cite{Thrun05} as well as ground truth will be shown using figures, which will provide an intuitive image of the good performance of the proposed method. 
     \item \textbf{Root mean square of errors of the trajectory}: $RMSE =  \sqrt{  \Sigma^{i_e}_{i=i_s}{ e(i)^2 }/{(i_e-i_s)} } $, where $i_s$ and $i_e$ are the start point and end point of the whole trajectory, and $e(i)$ the error at the time point $i$. $e(i) = ||l(i) - g(t)||_2$, where $l(i)$ and $g(t)$ are estimated locations and ground truth in 2D, respectively. 
     \item \textbf{Cumulative distribution function (CDF) of localisation errors}: The CDF of localisation errors demonstrates the frequency that a localisation error is less than or equal to a corresponding error.
     \item \textbf{Localisation error of the end point}: As the proposed localisation method is based on the trajectory, a good quality localisation system should not have a huge cumulative error in the end.

 \end{itemize}
 
\subsection{Experiment in Real Environment}
To evaluate the proposed system, the user walks on one floor in a house within 10 m $\times$ 5 m area. Figure \ref{fig:traj_gt} shows the ground truth of the trajectory. The situation is similar to walking in a house, where there is furniture, so that people can only walk in a certain area. The ground truth is obtained by manually adding loop closures to key points, such as narrow corridors where people have to walk by.  

In this experiment, the total trajectory path is 308.8 m. Figure \ref{fig:traj_imu} shows the trajectory only using IMU. Compared with Figure \ref{fig:traj_gt}, there are two observations: (1) the shape of the trajectory is similar to the ground truth; (2) the trajectory has accumulative bias compared with the ground truth. This confirms that it is feasible to use an IMU based PDR method for the indoor localisation system, and a secondary mechanism, such as the loop closure method, is required to calibrate the raw trajectory. After considering WiFi RTT based loop closures, false positives are introduced, i.e., due to the noise and multipath effect, distant points could be clustered falsely for further loop closure. When using the traditional pose graph SLAM \cite{Thrun05}, as shown in Figure \ref{fig:traj_noslam}, it cannot provide accurate localisation performance. Different from traditional SLAM, the robust pose graph SLAM, as discussed, uses various weights for loop closures during the optimisation process, so the false positives of loop closures can be given less weight. Figure \ref{fig:traj_slam} shows the trajectory using the proposed method, which clearly shows its good performance and outperforms the IMU based PDR method. The trajectory in Figure \ref{fig:traj_slam} also confirms that the robust SLAM has the capability to handle the false positives of loop closures and obvious enhancement compared with the trajectory using traditional SLAM shown in Figure \ref{fig:traj_noslam}.

When looking at RMSE of the trajectory, the use of IMU PDR can achieve 2.67 m. When using traditional SLAM, the RMSE of the trajectory is 5.95 m. When using the proposed method, the RMSE of the trajectory can reach 0.13 m, which has significantly been increased compared with the use of IMU and traditional SLAM. The good performance of the proposed method can be further confirmed by the localisation error of the end point. When using IMU based PDR, the localisation error of the end point is 9.84 m. The use of the traditional SLAM and our proposed method can achieve nearly 0 m. The good performance in this metric for the proposed method and traditional SLAM is due to loop closures near the end point. Although the use of traditional SLAM can achieve good performance in this metric, it can still be clearly seen in Figure \ref{fig:traj_noslam} that, for the most part of its estimate trajectory, the use of the traditional SLAM cannot achieve satisfactory localisation performance. 

Figure \ref{fig:cdf} shows the CDF of errors using different methods. It demonstrates that the 90th percentile tracking errors for the IMU based PDR, traditional SLAM and the proposed method are 9.12 m and 4.46 m, and 0.21 m, respectively. This means, when using the CDF of localisation errors as a metric, the proposed method can have 97.6\% and 95.2\% enhancement compared with the PDR and traditional SLAM, respectively.


\section{CONCLUSIONS}\label{sec:conclusion}

In this paper, we use WiFi RTT measurements as observations for loop closure. A robust pose graph SLAM system is applied on both types of constraints, i.e., raw trajectories from IMU measurements and loop closures from WiFi RTT measurements. An efficient clustering method for WiFi RTT measurements has been proposed for loop closure, which, different from traditional methods, does not rely on the knowledge of the locations of WiFi access points and accurate ranging results. The robust pose graph SLAM can also help handle false positives of loop closures, which provides a good indoor localisation performance. To evaluate our proposed method, we have conducted experiments in a real indoor environment, which can achieve 0.13 m accuracy and clearly shows the proposed method outperforms the IMU based PDR and the localisation using traditional SLAM by more than 90\%.











\bibliographystyle{IEEEtran}
\bibliography{sigproc}

\end{document}